\documentstyle[12pt]{article}
\begin{document}
\begin{center}
\begin{large}
\mbox{}\vspace{5mm}

{\bf THE LOCAL STABILITY OF ACCRETION DISKS WITH ADVECTION}\\
[5mm]
\end{large}
\end{center}
\begin{center}
Xue-Bing Wu$^{1,2}$ and Qi-Bin Li$^{1}$ \\[5mm]
 1. Beijing Astronomical Observatory, Chinese Academy of Sciences, Beijing
 100080, China\\
 2. Department of Physics, The University of Hong Kong, Pokfulam Road, Hong
 Kong\\
 email: wuxb@yac.bao.ac.cn; lqb@bao01.bao.ac.cn\\[8mm]
\end{center}
\noindent
{\bf Received:}\\[3mm]
\begin{center}
{\bf Abstract}
\end{center}

Based on the disscusion of the applicability of local approximation, the local
stability of accretion disks with advection is studied together
with the considerations of radial viscous force and thermal diffusion.
For a geometrically thin, radiative cooling dominated disk, the thermal
diffusion has nearly no effects on the thermal and viscous modes, which are
both stable if the disk is also optically thick, gas pressure dominated and are
both unstable if the disk is whether optically thick, radiation pressure
dominated or optically thin. The including of thermal diffusion, however, tends
to stabilize the acoustic modes which, if without advection, are unstable if
the disk is optically thick, radiation pressure dominated or optically thin,
and are stable if the disk is optically thick, gas pressure dominated. The
including of very little advection has significant effects on two acoustic
modes, which are no longer complex conjugates each other. Independent on
the optical depth, the instability of
the outward propagating
mode (O-mode) is enhanced and that of the inward propagating mode (I-mode) is
damped if the disk is gas pressure dominated, while the instability of O-mode
is damped and that of I-mode is enhanced if the disk is radiation pressure
dominated. For a geometrically slim, advection-dominated disk, both the thermal
and
viscous modes, as well as I-mode, are always stable if the disk is optically
thin. The
including of thermal diffusion tends to make these modes more stable. However,
the O-mode can become unstable when $q/m$ is very large ($q$ is the ratio of
advective
to viscous dissipated energy and $m$ the Mach number), even if the thermal
diffusion
is considered. On the other hand, if the advection-dominated disk is optically
thick, we found there is no self-consistent acoustic modes in our local
analyses.
The thermal diffusion has no effect on the stable viscous mode but has a
significant contribution to enhance the thermal instability.\\

\noindent
{\it Subject headings:} Accretion, accretion diks - instabilities\\[5mm]
\newpage
\begin{center} 1. Introduction \end{center}

    The stability of geometrically thin accretion disks has been extensively
studied after the construction of standard $\alpha$ model (Shakura \& Sunyaev
 1973). It has been found that the disk is thermally and viscously unstable if
it is optically thick and radiation pressure dominated (Pringle, Rees \&
Pacholczyk
1973;
Lightman \& Eardley
1976; Shakura \& Sunyaev 1976). Subsequent studies have found that the
optically
thick disk admits not only the thermal and viscous instabilities but also the
inertial-acoustic (or pulsational) instability (Kato 1978;
Blumenthal, Yang \& Lin 1984). If the geometrically thin disk is optically
thin, it has been also
found it is viscously stable but thermally unstable (Piran 1978). Those
instabilities are believed to be relevant
to some light variations observed in many systems such as cataclysmic
variables,
X-ray binaries and active galactic nucleus. For example, the thermal
instability
may account for the periodic outburst of dwarf nova (Osaki 1974) and the
inertial-acoustic instability may explain the observed QPO phenomena in
Galactic
 black hole candidates (Chen \& Taam 1995).

 In the standard $\alpha$ model, the viscous heating balances by radiative
cooling. However, if the radiative cooling is not efficient, the advection
will be not negligible. Particularly in an optically thin disk, the radiative
cooling rate is so low that most of the viscous generated energy is advected
radially. Recently, the accretion disk models with advection
have been studied when the disk is either optically thick or optically thin (
Abramowicz et al. 1988; Kato, Honma \& Matsumoto
1988; Narayan \& Popham 1993; Narayan \& Yi 1994, 1995a,b; Abramowicz et al.
1995; Chen et al. 1995; Chen 1995). The advection-dominated disk model has also
been successfully adopted to explain the observations of both low luminosity
and high luminosity systems (Narayan, Yi \& Mahadcvan 1995; Narayan, McClintock
 \& Yi 1996; Lasota et al 1996; Narayan 1996).

Although
some of the stability properties of accretion disks with advection has been
suggested in the previous research by analysing the disk structure, the
detailed
stability analysis has not yet been well done. From the $\dot{M}-\Sigma$
relation, the
stability properties of thermal and viscous modes in accretion disks
with advection can be obtained by analysing the $\dot{M}(\Sigma)$ slope and
comparing the cooling and heating rates near each equilibrium curve. In
particular,
the advection-dominated disks are
suggested to be both thermal and viscous stable whether the disks are optically
thin or optically thick (Chen at al 1995). However, such a stability analysis
can only be applied
to the long-wavelength perturbations (Chen 1996). In order to confirm the
previous results,
the detailed stability analyses of accretion
disks should be done by considering the perturbations to the time-dependent
equations. More recently, Kato, Abramowicz \& Chen (1996) performed an analytic
stability analysis to the advection-dominated disks by considering the local
perturbations. They found the optically
thick disks are still thermally unstable when the thermal diffusion is
considered,
 while
the optically thin disks are always stable. Furthermore, they thougth that in
the
case of
an advection
-dominated disk, the variations of angular momentum and of surface density
associated with the perturbations lead to thermal instability, which is quite
different from that in a radiative cooling dominated disk where there is no
appreaciable surface density change (Pringle 1976).

Stimulated by the above research, we have performed a detailed study to the
local
stability of accretion disks with advection. The influences of thermal
diffusion
on the disk instability were also considered. Different from the study of Kato
et
al. (1996), we discussed not only the stability of thermal mode, but also the
viscous and inertial-acoustic modes. Moreover, a general dispersion relation,
 which is suitable for the stability analyse of accretion disks with
different disk structures, was obtained. In Section 2 we give the basic
equations and a discussion about the validity of local  approximation. In
Section
3 we derive the perturbed equations and the dispersion relation. The stability
properties of optically thin and optically thick disks are presented
respectively
in Section 4 and Section 5. Finally in Section 6, a brief discussion about our
results is given.\\[4mm]

\noindent
\begin{center} 2. Basic equations and local approximation
\end{center}

We consider an axisymmetric and non-self-gravitating accretion disk. The
effects of general relativity are introduced by the pseudo-Newtonian
potential (Paczynski \& Wiita 1980), $\Psi=-GM/(R-r_{g})$, where $M$ is the
mass of
central object,
$R=(r^{2}+z^{2})^{1/2}$ and $ r_{g}=2GM/c^{2}$. Adopting
a cylindrical system of coordinates ($r,\varphi, z$) which is centered on the
central object, the vertical integrated time-dependent equations describing
accretion flow can be written as:
\begin{equation}
\frac{\partial{\Sigma}}{\partial{t}}+\frac{1}{r}\frac{\partial}{\partial{r}}
(r\Sigma V_{r})=0,
\end{equation}
\begin{equation}
\Sigma\frac{\partial{V_{r}}}{\partial{t}}+\Sigma V_{r}\frac{\partial{V_{r}}}
{\partial{r}}-\Sigma(\Omega^{2}-{\Omega_{k}}^{2})r=-2\frac{\partial{(Hp)}}{\partial
{r}}+F_{\nu},
\end{equation}
\begin{equation}
\Sigma r^{3}\frac{\partial{\Omega}}{\partial{t}}+\Sigma rV_{r}\frac{\partial{
(r^{2}\Omega)}}
{\partial{r}}=\frac{\partial}{\partial{r}}(\Sigma \nu r^{3}
\frac{\partial{\Omega}}
{\partial{r}}),
\end{equation}
\begin{equation}
C_{v}[\Sigma\frac{\partial{T}}{\partial{t}}+\Sigma V_{r}\frac{\partial{T}}{
\partial{r}}-(\Gamma_{3}-1)T(\frac{\partial{\Sigma}}{\partial{t}}+V_{r}\frac
{\partial{\Sigma}}{\partial{r}})]=\Sigma \nu (r\frac{\partial{\Omega}}{\partial
{r}})^{2}-Q_{-}+Q_{t},
\end{equation}
where $ V_{r}$, $\Omega$ are the radial velocity and angular velocity, $p$, $T$
and
$\Sigma$ are the total pressure, temperature and surface density, $C_{v}$ and
$\Gamma_{3}$ are the heat capacity per unit mass and a quantity associated with
$\beta$, the ratio of gas to total pressure. The total pressure $p$ is the sum
of
gas and radiation pressures given by $p=\Re \rho T +aT^{4}$, where $\Re$ is the
gas
constant
and $a$ the radiation constant. $\Omega_{k}$ is the Keplerian angular
velocity, given by ${\Omega_{k}}^{2}=(\frac{\partial{\Psi}}{r
\partial{r}})_{z=0}
$. $F_{\nu}$ is the radial viscous force, which is often neglected in
geometrically
thin accretion disks but is perhaps not negligible in accretion disks with
advection. It is given by (Papaloizou \& Stanley 1986)
\begin{equation}
F_{\nu}=\frac{\partial}{\partial{r}}[\frac{1}{2} \frac{\nu_{r}\Sigma}{r} \frac
{\partial{(rV_{r})}}{\partial{r}}]-\frac{2V_{r}}{r}\frac{\partial{(\nu_{r}\Sigma)}}
{\partial{r}},
\end{equation}
where $\nu_{r}$ is the kinematic viscosity acting in the radial direction. In
this
paper we take $\nu_{r}=\nu$, where $\nu$ is the viscosity acting in the
azimuthal
direction and is expressed as the standard $\alpha$ prescription (Shakura \&
Sunyaev 1973), $\nu=\alpha c_{s}H$. $c_{s}$ is the local sound speed defined by
$c_{s}=p/\rho$, where $\rho$ is the density. $H$ is the disk height given by
$H=c_{s}/\Omega_{k}$. $Q_{-}$ at the right side of Eq. (4) represents the
radiative
cooling. For a optically thick disk, we take it as
\begin{equation}
Q_{-}=\frac{8acT^{4}}{3\kappa\Sigma},
\end{equation}
where $\kappa$ is the opacity given by electron
scattering: $\kappa_{es}=0.34$. For an optically thin disk, we assume $Q_{-}$
is provided by thermal
bremsstrahlung with emissivity
($erg s^{-1} cm^{-2}$)
\begin{equation}
Q_{-}=1.24\times 10^{21} H\rho^{2}T^{1/2}A,
\end{equation}
where $A\ge 1$ is the Compton enhancement factor. $Q_{t}$ at the right side of
Eq.
(4) represents the thermal diffusion defined as $Q_{t}=\nabla\cdot(K
\nabla T)$, where $K$ is the vertical integrated thermal conductivity
given by $K=\Sigma C_{v}\nu =\alpha f\Omega_{k}H^{3}p/T$, where $f=3(8-7\beta)
f_{*}$ and $f_{*}$ is of the order of unity (Kato et al.
1996).

 In this paper we will consider the local perturbations to the accretion disks.
The radial perturbations of $V_{r}, \Omega, \Sigma$ and $T$ are of the form $(
\delta V_{r}, \delta \Omega, \delta \Sigma, \delta T)\sim e^{i(\omega t-kr)}$,
where $k$ is the perturbation wavenumber defined by $k=2\pi/\lambda$, $\lambda$
is
the perturbation wavelength.  The local approximation means $\lambda < r$ and
the validity
of the vertically integrated equations requires $kV_{r} < \Omega_{k}$. Since
$V_{r} \sim \alpha {c_{s}}^{2}/\Omega_{k}$, the requirements above can be
written
as
\begin{equation}
\frac{r}{H}> \frac{\lambda}{H} > 2\pi \alpha \frac{H}{r}.
\end{equation}
We can see clearly that this inequality is well satisfied for a geometrically
thin
accretion disk even if we set $\lambda/H$ in a wide range, such as from 1 to
100. However, for a geometrically slim disk, where $H/r \le 1$, it can be
satisfied
only when $\alpha$ is sufficiently small. The range of $\lambda/H$ also moves
to
the smaller value such as from 0.06 to 1 if $\alpha$ is about 0.01. In
addition,
the validity of vertical integrated equations also requires the growth rates of
unstable modes are less than the angular velocity (Kato et al. 1996). In order
to
get reasonable and self-consistent results, we present
our discussions in this paper following all these restrictions.\\

\noindent
\begin{center}
3. Perturbed equations and dispersion relation
\end{center}

Considering the radial perturbations to $V_{r}, \Omega, \Sigma$, $T$ and the
local approximation, the perturbed equations corresponding to Eqs.(1)-(4) can
be
written as followings after a lengthy deduction.
\begin{equation}
\tilde{\sigma}\frac{\delta\Sigma}{\Sigma}-i \frac{\epsilon}{\tilde{H}}
\frac{\delta
V_{r}}{\Omega_{k}r}=0,
\end{equation}
\begin{equation}
-i \epsilon \tilde{H} \frac{\delta \Sigma}{\Sigma}+(\tilde{\sigma}+\frac{4}{3}
\alpha \epsilon^{2})\frac{\delta V_{r}}{\Omega_{k}
r}-2\tilde{\Omega}\frac{\delta
 \Omega}{\Omega_{k}}-i x_{1}\epsilon \tilde{H} \frac{\delta T}{T}=0,
\end{equation}
\begin{equation}
 i x_{2}\alpha \epsilon\tilde{H}g\frac{\delta
\Sigma}{\Sigma}+\frac{{\tilde{\chi}}^{2}}
 {2\tilde{\Omega}}\frac{\delta V_{r}}{\Omega_{k}
r}+(\tilde{\sigma}+\alpha\epsilon^{2}
 )\frac{\delta \Omega}{\Omega_{k}}+ix_{3}\alpha\epsilon\tilde{H}g\frac{\delta
T}
 {T}=0,
\end{equation}
\begin{equation}
 -(y_{1}\tilde{\sigma}+\alpha y_{2}
g^{2})\frac{\delta\Sigma}{\Sigma}-\frac{\alpha q
 g^{2}}{m\tilde{H}}\frac{\delta V_{r}}{\Omega_{k} r}+\frac{2i\alpha\epsilon g}
 {\tilde{H}}\frac{\delta{\Omega}}{\Omega_{k}}+(y_{4}\tilde{\sigma}-\alpha x_{2}
 g^{2}+\alpha f \epsilon^{2})\frac{\delta T}{T}=0,
\end{equation}
where $\tilde{\sigma}=\sigma/\Omega_{k}$, and $\sigma=i(\omega-kV_{r})$.
$\tilde
{\Omega}=\Omega/\Omega_{k}$, $\tilde{H}=H/r$, and $\epsilon=kH$.
$g=\frac{{\tilde{\chi
}}^{2}}{2\tilde{\Omega}}-2\tilde{\Omega}$, and $\tilde{\chi}=\chi/\Omega_{k}$
where $\chi$ is the epicyclic frequency defined by $ \chi^{2}=2\Omega(2\Omega
+r\frac {\partial \Omega}{\partial r})$. $m$ is the Mach number defined by
$m=\mid V_{r}\mid/c_{s}$. $q$ is the ratio of advective energy to
viscous dissipated energy, namely
\begin{equation}
C_{v}[\Sigma V_{r}\frac{\partial{T}}{
\partial{r}}-(\Gamma_{3}-1)T(\frac{\partial{\Sigma}}{\partial{t}}+V_{r}\frac
{\partial{\Sigma}}{\partial{r}})]=q \Sigma \nu
(r\frac{\partial{\Omega}}{\partial
{r}})^{2}.
\end{equation}
If the disk is radiative cooling dominated, $q$ is nearly zero and if it is
advection-dominated, $q$ is nearly 1. The values of $x_{1}$, $x_{2}$, $x_{3}$,
$y_{1}$,
$y_{2}$, $y_{3}$ and $y_{4}$ in the perturbed equations depend on the structure
properties of accretion disks. For a gas pressure dominated disk, $x_{1}=x_{2}
=x_{3}=1$, and $y_{1}=1$, $y_{2}=1-2q$, $y_{3}=\frac{3}{2}$, $y_{4}=-1$ if it
is
optically
thin while $y_{1}=1$, $y_{2}=2-q$, $y_{3}=\frac{3}{2}$, $y_{4}=3-4q$ if it is
optically
thick. For a radiation pressure dominated disk, $x_{1}=-8$, $x_{2}=-1$,
$x_{3}=8,$ and
$y_{1}=4$, $y_{2}=-q$, $y_{3}=12$, $y_{4}=-4(1+q)$ if it is optically thick.

By setting the determinants of the coefficients in above perturbed equations
to zero, we get a dispersion relation:
\begin{equation}
a_{1}{\tilde{\sigma}}^{4}+a_{2}{\tilde{\sigma}}^{3}+a_{3}{\tilde{\sigma}}^{2}+
a_{4}\tilde{\sigma}+a_{5}=0,
\end{equation}
where $a_{i} (i=1,...,5)$ is the coefficients given by
$$
a_{1}=y_{3},$$
$$
a_{2}=\alpha[\epsilon^{2}(f+\frac{7}{3}y_{3})+y_{4}g^{2}],$$
$$
a_{3}=\alpha\epsilon g^{2}[\alpha\epsilon(\frac{7}{3}y_{4}+2x_{3})-ix_{1}
\frac{q}{m}]+\frac{1}{3}(\alpha\epsilon^{2})^{2}(4y_{3}+7f)+\epsilon^{2}(
y_{3}+x_{1}y_{1})+y_{3}{\tilde{\chi}}^{2},$$
$$
a_{4}=2ix_{3}\alpha^{2}\tilde{\Omega}\epsilon g^{3}\frac{q}{m}+\alpha g^{2}[
\frac{4}{3}y_{4}(\alpha\epsilon^{2})^{2}-ix_{1}\alpha\epsilon^{3}\frac{q}{m}
+\frac{8}{3}x_{3}(\alpha\epsilon^{2})^{2}+y_{4}{\tilde{\chi}}^{2}+(y_{4}+x_{1}y_{2})
\epsilon^{2}]+$$
$$\alpha\epsilon^{2} g[2x_{1}\frac{{\tilde{\chi}}^{2}}{2\tilde
{\Omega}}-2\tilde{\Omega}(y_{1}x_{3}+x_{2}y_{3})]+\alpha\epsilon^{2}[\frac{4}{3}
f(\alpha\epsilon^{2})^{2}+f{\tilde{\chi}}^{2}+\epsilon^{2}(y_{3}+f+x_{1}y_{1})],
$$
$$
a_{5}=(\alpha\epsilon)^{2}[-2\tilde{\Omega}g^{3}(x_{3}y_{2}+x_{2}y_{4})+\epsilon^{2}
g^{2}(-2x_{1}x_{2}+x_{1}y_{2}+2x_{3})+\epsilon^2 g(y_{4}-2x_{2}f\tilde{\Omega})
+\epsilon^{4}f].
$$

The stability properties of two inertial-acoustic modes, thermal and viscous
modes
can be obtained by analyzing the four kinds of solutions of the dispersion
relation.
The real parts of these solutions correspond to the growth rates of the
perturbation modes and the imaginary parts correspond to their propagating
properties. In following
two sections we will numerically solve the dispersion relation according to the
different
disk structures. The stability of advection-dominated disks will be analyzed
by assuming
$q \rightarrow 1$. We note that the influence of radial viscous force on the
disk
stability
has been investigated by some authors (Papaloizou \& Stanley 1986; Wu, Yang \&
Yang 1994). Although it is included in our present study,  more attentions in
following sections will be paid on the effects of advection and thermal
diffusion. In addition, we note that $Re(\tilde{\sigma})$ is nearly in
proportional
to $\alpha$ in all cases. Thus, the influence of different viscosity parameters
on the disk stability will not be detailed studied.\\[4mm]

\noindent
\begin{center}
4. Stability of optically thin disks
\end{center}

Previous results show that an optically thin disk is viscously stable but
thermal
unstable if it is dominated by local radiative cooling, and is both viscously
and thermally stable if it is advection-dominated. In this section, we discuss
its stability according to the different contribution of advection. Some
parameters
in the dispersion relation, such as $ x_{i}$ and $y_{i}$, are taken as the
values
given in Section 3
for an optically thin, gas pressure
dominated disk. Others are dependent on the detailed disk
structure or are set as the variables in following cases.

(a) {\it Geometrically thin disk without advection}. It is purely radiative
cooling
by thermal bremsstrahlung that balances the viscous dissipation, then $q=0$.
We take $\tilde{\Omega}=\tilde{\chi}=1$,
$\alpha=0.01, m=0.01$.  $\lambda/H$ is set
from 1 to 80 for a geometrically thin disk according to the local restrictions
( Eq. (8)). By solving the dispersion
relation, we get the results shown in Fig.1. In the long wavelength limit,
 the thermal mode is unstable and the
viscous mode is marginally stable, which are in agreement with the previous
results. However, when $\lambda/H< 10$, the thermal and viscous modes are
stable and the acoustic instability becomes important. The thermal diffusion
has nearly no effect on the disk stability in the long wavelength limit, but
it  stabilizes the acoustic modes and thermal mode significantly when
the perturbation wavelength is shorter than $20H$.

(b) {\it Geometrically thin disk with very little advection}. The disk in
dominated
by radiative cooling but with very little advection, then we assume $q=0.01$.
Other
parameters are the same as in case (a). In Fig. 2, $m=0.01$, we see the thermal
and viscous modes are nearly the same as in Fig. 1 but the stability properties
of acoustic modes are quite differen. If even very little advection is present,
two acoustic modes are no longer complex conjugates as in the case without
advection. The Inward propagating mode
(I-mode) becomes stable while the outward propagating one (O-mode) becomes more
unstable. In addition, the inclusion of thermal diffusion has the same effect
as in case (a), which  stabilizes the  acoustic instability and thermal
instability significantly especially in the short wavelength case. Moreover, we
note the contribution of advection is associated with $q/m$, which
can be clearly seen in the dispersion relation. It has nearly no effect on the
thermal and viscous modes but has significant effects on two acoustic modes. In
Fig. 3, we see the effect of advection increases as the increase of the value
of $q/m$. If $q/m< 0.1$, however, this  effect becomes negligible and
the disk stability is nearly the same as in case (a).

(c) {\it Geometrically slim and advection-dominated disk}. In this case, the
viscous
dissipation is primarily balanced by advection and we take $q=0.99$, $\alpha=
0.001, H/r=0.6$. According to the local restrictions, $\lambda/H$ is set from
0.02
to 2. If we choose $\tilde{\Omega}=\tilde{\chi}=1$ and $m=0.1$, the solutions
of the dispersion relation are shown in Fig. 4. We can clearly see that all
four
modes are all stable. The thermal diffusion tends to make the disk more stable.
The O-mode and I-mode are not complex conjugates but the departure becomes less
as the decrease of perturbation wavelength. Moreover, from Fig. 5 we can see,
the departure of two acoustic modes becomes more less if $q/m$ decreases. Such
a departure is negligible if $q/m<0.3$. But if $q/m>100$, the O-mode can become
unstable when $\lambda/H > 0.8$, even if the thermal diffusion is considered.
In addition, we note that the optically
thin, advection dominated disk is probably sub-Keplerian (Narayan
\& Yi 1994). Fig. 6 shows the solutions with  $\tilde{\Omega}=\tilde{\chi}=
0.01$. In comparison with the case in Fig. 4, we note the viscous mode now
becomes slightly
unstable ($Re(\tilde{\sigma})/\alpha \sim 0.1$) if without thermal diffusion.
However, no viscous instability exits in a sub-Keplerian disk if the thermal
diffusion is considered.
The O-mode can becomes unstable only when $q/m$ is very large, such as
$q/m>700$.
The change of $q/m$ has significant effects on two acoustic modes, but has
nearly no effects on the thermal and viscous modes. \\[4mm]

\noindent
\begin{center}
5. Stability of optically thick disks
\end{center}

It is well known that an optically thick disk is thermally and viscously
unstable if it is dominated by radiation pressure, and is thermally and
viscously
stable if it is dominated by gas pressure. If it is advection dominated, recent
research suggested it is also both thermally and viscously stable. The local
stability analyses of an optically thick disk without advection and thermal
diffusion
have been performed by Blumenthal et al. (1984) and Wu at al. (1995a, b). In
this
section,
we will re-discuss
its stability in detail according to the different disk structures and the
contributions
of
advection and thermal diffusion. Some parameters
in the dispersion relation, such as $ x_{i}$ and $y_{i}$, are taken as the
values given
in Section 3
for an optically thick disk. $\tilde{\Omega}=\tilde{\chi}=1$ is adopted through
this section. Others are dependent on the detailed disk
structure or are set as the variables .

(a) {\it Geometrically thin, gas pressure dominated disk without advection}. We
take
$q=0$, $\alpha=0.01$, $m=0.001$. $\lambda/H$ is set from 1 to 80. Fig. 7 shows
the
solutions. We see the thermal and viscous modes are stable but the acoustic
modes are slightly unstable when $\lambda/H>7$, if without the thermal
diffusion.
However, the acoustic modes become stable if the thermal diffusion is
considered.
In addition, we note that the thermal diffusion has nearly no effects on the
thermal and viscous
modes, especially when $\lambda/H>10$.

(b) {\it Geometrically thin, gas pressure dominated disk with very little
advection}.
The disk is dominated by radiative cooling, so we take $q=0.01$. Other
variables
have the same values as in case (a). From Fig. 8, we can clearly see that two
acoustic modes now depart from each other due to the very little advection.
The O-mode becomes more unstable while the I-mode becomes stable. The inclusion
of thermal diffusion does not change the instability of O-mode, although it
decreases the growth rate of O-mode. The departure of two acoustic modes
will become less as the decrease of $q/m$. The inclusion of thermal diffusion
and the change of $q/m$, however, have nearly no effects on the thermal and
viscous modes, which are always stable.

(c) {\it Geometrically thin, radiation pressure dominated disk without
advection}.
In this case, we take $q=0, \alpha=0.01, m=0.01$ and $\lambda/H$ from 1 to 80.
Fig. 9 shows the similar result as those in Blumenthal et al (1984) and Wu
et al. (1995a). The inclusion of thermal diffusion decreases the growth rate of
acoustic modes but has very little effects on the thermal and viscous modes.
The thermal and viscous instabilities are dominant for long wavelength
perturbations
but the acoustic instability is dominant when $\lambda/H < 20$.

(d) {\it Geometrically thin, radiation pressure dominated disk with very little
advection}. Fig. 10 shows the case similar as in Fig. 9 but with $q=0.01$. In
comparison with case (c), we see that the inclusion of very little advection
leads to a singificant departure of two acoustic modes but has nearly no
effects
on the
thermal and viscous instabilities. The thermal diffusion tends
to stabilize the acoustic modes but does not alter the instability of
I-mode. The influence of advection term with different values of $q/m$ on the
acoustic modes is shown
in Fig. 11. The decrease of
$q/m$ has significant effects to lessen both the instability
of I-mode and the stability of O-mode by reducing the departure of two
acoustic modes,.

(e) {\it Geometrically slim, radiation pressure and advection dominated disk}.
In this
case, we take $q=0.99$, $\alpha=0.001$, $m=0.1$ and $\lambda/H$ from 0.02 to 2
for a slim disk according
to the local restrictions. We note that there are no self-consistent solutions
corresponding to two acoustic modes. Fig. 12 shows the results for thermal and
viscous modes. We can clearly see that the thermal diffusion has a significant
role to enhance the thermal instability. The growth rate of thermal instability
increases quickly as the decrease of perturbation wavelength. If without
thermal
diffusion, the thermal mode is only slightly unstable. The viscous mode,
however,
is always stable and does not change even if the thermal diffusion is included.
In
addition, we
have also investigated the effects of the changes
of $ q/m$, $\alpha$ and the rotation law on the thermal and viscous modes.
Those
effects were found negligible. The reason why the acoustic modes disappears in
our local analyses can be seen by investigating the dispersion relation (Eq.
14). In the limit when $\alpha \sim 0$, the dispersion relation becomes
\begin{equation}
y_{3}{\tilde{\sigma}}^{4}+[y_{3}{\tilde{\chi}}^{2}+(y_{3}+x_{1}y_{1})\epsilon^{2}]
{\tilde{\sigma}}^{2}=0.
\end{equation}
Except two trivial solutions, other two solutions of above equation represent
two
propagating acoustic modes, which are described by
\begin{equation}
{\tilde{\sigma}}=\pm i[{\tilde{\chi}}^{2}+(1+\frac{x_{1}y_{1}}{y_{3}})\epsilon
^{2}]^{1/2},
\end{equation}
where the positive imaginary part corresponds to the circle frequency of O-mode
while the negative one corresponds to that of I-mode.
The existence of self-consistent propagating acoustic modes requires the term
in
the square
bracket of Eq. (16) is positive. For an optically thick, radiation pressure
dominated disk, $x_{1}=-8$, $y_{1}=
4$ and $y_{3}=12$. The requirement
above means $\lambda/H>\sqrt{\frac{5}{3}}\frac{2\pi}{\tilde{\chi}}$. If we
choose ${\tilde{\chi}}={\tilde{\Omega}}=1$, the acoustic modes exist only when
$\lambda/H>8$, which is satisfied for a geometrically thin disk but not for a
geometrically slim disk when a local analyses is presented. However, for a
gas pressure dominated disk, where $x_{1}=y_{1}=1$ and $y_{3}=\frac{3}{2}$,
we can see from Eq. (16) that the acoustic modes always exist independent of
the opacity and the geometry of the disk.\\[4mm]

\noindent
\begin{center}
6. Discussion
\end{center}

We have performed detailed analyses to the local stability of accretion disks
with advection. We found that for a geometrically thin, radiative cooling
dominated
disk, the presence of even
very little advection has significant effects on the acoustic modes.
Furthermore,
those effects depend on what kinds of pressure is dominated in the disk. If the
disk
is gas pressure dominated, the presence of very little advection enhances the
instability of O-mode and damps that of I-mode. But if it is radiation pressure
dominated, the I-mode will become more unstable while the O-mode tends to
become
stable. Those effects are independent on the optical depth of the disk. The
presence of very little advection has nearly no effect on the thermal and
viscous modes. They are unstable in an optically thin disk
and in an optically thick, radiation pressure dominated disk, but are stable
in an optically thick, gas pressure dominated disk. These results
are well agreement with those obtained in previous studies on the stability
of a geometrically thin disk.
In addition, we note that the inclusion of thermal diffusion has a significant
effect to stabilize the acoustic modes in the short perturbation wavelength
case,
but
has nearly no effects on the thermal and viscous modes especially in the long
perturbation wavelength case.
 For a geometrically slim, advection-dominated disk, it is in general stable
if it is optically thin.  The
viscous mode is slightly unstable in a sub-keplerian disk but will also become
stable if the thermal diffusion is included. Only the O-mode can become
unstable when $q/m>100$ even if the thermal diffusion is considered. If the
disk
is optically thick and advection-dominated, we found no self-consistent
acoustic
modes exist in our local analysis.
The thermal mode is slightly unstable if without thermal diffusion, and it
becomes
much more unstable if the thermal diffusion is considered. The growth rate of
thermal
mode increases rapidly as the decrease of perturbation wavelength. It means the
thermal
instability is very important for an optically thick, advection-dominated disk,
which
was previously suggested to be thermally stable.
It has been pointed out that the thermal instability of an optically thick,
advection-dominated
disk is due to a large density
change associated with a small pressure change (Kato et al. 1996), which is
different from the case of a geometrically thin disk where there is no
appreaciable
change of surface density.

The significant effects of very little advection on the acoustic modes is not
surprised. The departure of O-mode and I-mode have been found by Chen \& Taam
(1993) and Wu \& Yang (1994) when a casually limited viscosity is adopted.
Together with the results in the present study, we think such a departure is
resluted from the effects associated with the radial velocity. Due to the
different
disk structures, the influences
of advection term on the acoustic modes in the disk with different kinds of
dominant pressure are also different. In addition, we note the importance of
thermal diffusion on the stability of advection-dominated disk
has been pointed out by Kato et al. (1996). Our results prove that the thermal
diffusion has a significant effects to damp the instability of an optically
thin,
advection dominated disk but
enhance the thermal instability of an optically thick, advection-dominated
disk.
We can also see from Eq. (13)
that the thermal diffusion terms is in proportional to $(kH)^{2}$, which can
ne neglected for a geometrically thin disk but can not be neglected for a
geometrically
slim disk where $kH \geq 1$. Because the advection-dominated disk is usually
not
geometrically thin,
the thermal diffusion must be included in a stability analysis. In addition, we
note that the ratio of thermal discussion to viscous dissipation, $Q_{t}/\Sigma
\nu(r\frac{\partial\Omega}{\partial r})$, is in proportional to $(H/r)^{2}$. It
also means the thermal diffusion is not negligible when the structure of
advection-
dominated disk is constructed. From the well agreement of our results with some
stability properties obtained from the analyses of $\dot{M}(\Sigma)$ slope, we
think the structure of an advection-dominated disk may be a little different
if the thermal diffusion is included. Thus, a detailed calculation of disk
structure
with thermal diffusion is expected.

Finally, we would like to mention that our local stability analyses are
performed
under some simplifications. Some physics processes neglected in our present
study may
not be negligible in
some cases. For example, the two temperature effects in an optically thin disk
may be very important if the disk temperature exceeds $10^{9}$K (Shapiro,
Lightman
 \& Eardley 1976). The viscous
dissipation in radial direction is also neglected in this paper. In addition,
we must note that the local stability may not be the same as that obtained from
a
global analysis. Moreover,
the physics reason for the thermal instability in an optically thick, advection
-dominated disk, which has been suggested by Kato et al. (1996), need to be
confirmed
in another detailed future work. \\[3mm]

XBW thanks Shoji Kato, Xingming Chen for helpful disccusions on the thermal
diffusion and advection. He is also much grateful to K. S. Cheng and the
University
of Hong
Kong for kind hospitality. The work is partly supported by the National
Climbing
Program on Fundamenthal Research of China.\\[4mm]

\noindent
{\bf References}\\[3mm]
\noindent
Abramowicz, M.A., Czerny, B., Lasota, J.P., Szuszkiewicz, E. 1988, ApJ, 332,
646\\
Abramowicz, M.A., Chen, X., Kato, S., Lasota, J.-P., Ragev, O. 1995, ApJ, 438,
L37\\
Blumenthal, G.R., Yang, L.T., Lin, D.N.C. 1984, ApJ, 287, 774\\
Chen, X. 1995, MNRAS, 275, 641\\
Chen, X. 1996, in ``Basic Physics  of Accretion Disks" ed. by Kato S. et al.,
Gondon and Breach Science Publishers, in press\\
Chen, X., Abramowicz, M.A., Lasota, J.-P., Narayan, R., Yi, I. 1995, ApJ, 443,
L61\\
Chen, X., Taam, R.E. 1993, ApJ, 412, 254\\
Chen, X., Taam, R.E. 1995, ApJ, 452, 379\\
Kato, S. 1978, MNRAS, 185, 629\\
Kato, S., Abramowicz, M.A., Chen, X. 1996, PASJ, in press\\
Kato, S., Honma, F., Matsumoto, R. 1988, MNRAS, 231, 37\\
Lightman, A., Eardely, D. 1974, ApJ, 187, L1\\
Narayan, R. 1996, ApJ, in press\\
Narayan, R., McClintock, J.E., Yi, I. 1996, ApJ, in press\\
Narayan, R., Popham, R. 1993, Nature, 362, 820\\
Narayan, R., Yi, I. 1994, ApJ, 428, L13\\
Narayan, R., Yi, I. 1995a, ApJ, 444, 231\\
Narayan, R., Yi, I. 1995b, ApJ, in press\\
Narayan, R., Yi, I., Mahadevan, R. 1995, Nature, 374, 623\\
Lasota, J.-P., Abromowicz, M., Chen, X., Krolik, J., Narayan, R., Yi, I. 1996,
ApJ, in press\\
Osaki, Y. 1974, PASJ, 26, 429\\
Pacy\'nski, B., Wiita, P.J., 1980, A\&A, 88, 23\\
Papaloizou, J.C.B., Stanley, G.Q.G. 1986, MNRAS, 220, 593\\
Piran, T. 1978, ApJ, 221, 652\\
Pringle, J.E., Rees, M.J., Pacholczyk, A.G. 1973, A\&A, 29, 179\\
Shakura, N.I., Sunyaev, R.A. 1973, A\&A, 24, 337\\
Shakura, N.I., Sunyaev, R.A. 1976, MNRAS, 175, 613\\
Shapiro, S.L., Lightman, A.P., Eardley, D.N. 1976, ApJ, 204, 187\\
Wu, X.B., Li, Q.B., Zhao, Y.H., Yang, L.T. 1995a, ApJ, 442, 736\\
Wu, X.B., Li, Q.B., Zhao, Y.H., Yang, L.T. 1995b, ApJ, 442, 743\\
Wu, X.B., Yang, L.T. 1994, ApJ, 432, 672\\
Wu, X.B., Yang, L.T., Yang, P.B. 1994, MNRAS, 270, 465\\

\newpage
\begin{center}
{\bf FIGURE CAPTIONS}\\
\end{center}
\noindent
{\bf Figure 1.} The stability of an optically thin disk without advection.
The solid, long dashed and shord dashed lines correspond to acoustic modes,
thermal and viscous modes respectively. Those lines with star centered
represent
the modes in the case where the thermal diffusion is considered.\\[3mm]
{\bf Figure 2.} The stability of an optically thin disk with very little
advection. The solid and more short dashed lines correspond to the I-mode
and O-mode. Others have the same meanings as in Fig. 1.\\[3mm]
{\bf Figure 3.} The influences of different $q/m$ on the stability of acoustic
modes
in an optically thin disk with very little advection. The lines without any
symbol
centered, and those with open circle and with star centered correpond to $q/m=
1$, 5 and 0.1 respectively.\\[3mm]
{\bf Figure 4.} The stability of an optically thin, advection-dominated disk.
The solid, more short dashed, long dashed and short dashed lines correspond
to I-mode, O-mode, thermal mode and viscous modes respectively. The lines with
star centered correpond to those modes in the case with thermal diffusion
considered.\\[3mm]
{\bf Figure 5.} The influences of different $q/m$ on the stability of acoustic
modes in an optically thin, advection-dominated disk. The lines without any
symbol centered, and those with open circle and with star centered correpond
to $q/m=$10, 0.33 and 100 respectively.\\[3mm]
{\bf Figure 6.} The same as Fig. 4 but for a sub-Keplerian disk where $\tilde
{\Omega}=\tilde{\chi}=0.01$. The solid and more short dashed lines with open
circle centered represent respectly the I-mode and O-mode in the case with
$q/m=
1000$.\\[3mm]
{\bf Figure 7.} The stability of an optically thick, gas pressure dominated
disk
without advection. The solid, long dashed and short dashed lines correspond to
acoustic modes, thermal and viscous modes respectively. Those lines with star
centered represent the case with thermal diffusion.\\[3mm]
{\bf Figure 8.} The stability of an optically thick, gas pressure dominated
disk with very little advection. The solid and more short dashed lines with
open circle centered correpond to I-mode and O-mode in the case with $q/m=1$.
Other lines have the same meanings as those in Fig. 4 and with $q=0.01$, $m
=0.001$.\\[3mm]
{\bf Figure 9} The stability of an optically thick, radiation pressure
dominated
disk without advection. The lines have the same meanings as in Fig. 7.\\[3mm]
{\bf Figure 10} The stability of an optically thick, radiation pressure
dominated
disk with very little advection. The lines have the same meanings as in Fig.
7.\\[3mm]
{\bf Figure 11} The influences of different $q/m$ on the stability of acoustic
modes in an optically thick, radiation pressure dominated disk with very little
advection. The lines without any
symbol centered, and those with open circle and with star centered correpond
to $q/m=1$, 0.2 and 2 respectively.\\[3mm]
{\bf Figure 12.} The stability of an optically thick, radiation pressure and
advection dominated disk. The long dashed and short dashed lines correspond to
the thermal mode and viscous mode. The lines with star centered represent the
case
with thermal diffusion.

\end{document}